# A new Watermarking Technique for Medical Image using Hierarchical Encryption

Med Karim ABDMOULEH, Ali KHALFALLAH and Med Salim BOUHLEL

Research Unit: Sciences and Technologies of Image and Telecommunications
Higher Institute of Biotechnology
Sfax, Tunisia

**Abstract**
In recent years, characterized by the innovation of technology and the digital revolution, the field of media has become important. The transfer and exchange of multimedia data and duplication have become major concerns of researchers. Consequently, protecting copyrights and ensuring service safety is needed. Cryptography has a specific role, is to protect secret files against unauthorized access. In this paper, a hierarchical cryptosystem algorithm based on Logistic Map chaotic systems is proposed. The results show that the proposed method improves the security of the image. Experimental results on a database of 200 medical images show that the proposed method significantly gives better results.

## 1. Introduction

Image Encryption is a wide area of research and cryptography presents the basic science to secure information [1-3]. Encryption basically deals with converting data or information from their original form to some other form that changes information. The protection of the image data against unauthorized access is important. Encryption is also employed to increase the data security. The Encrypted Image is secure from any kind cryptanalysis.

Chaotic systems have been used in modern cryptosystem. Chaos has many important properties, such as the sensitive dependence on initial conditions and system parameters, pseudorandom property, non periodicity and topological transitivity, etc. These proprieties are very important in cryptography [4-8].

To resist to statistic and entropy attacks [1], we choose a "Logistic Map" (LM) function [9]. Indeed, the LM function ensures the properties of chaotic systems, such as sensitivity to initial values and control parameters.

In [10], authors improve the properties of confusion and diffusion in terms of discrete exponential chaotic maps, and design a key scheme to resist to statistic attack and differential attack attacks. In [11], Abdmouleh et al. propose a new encryption scheme for Magnetic Resonance Imaging (MRI), using the chaos theory to define a dynamic chaotic Look-Up Table (LUT). In [12], authors propose a new scheme for image encryption based on the use of a chaotic map with large key space and Engle Continued Fractions (ECF) map. In [13], Tabash et al. propose a new approach for image encryption based on three chaotic logistic maps and a multi-pseudo random block permutation.

The remainder of this paper is organized as follows. Section 2 introduces the architecture and implementation of the proposed algorithm. Section 3 illustrates a number of analyses of the algorithm performance. The security of our proposed method against known plaintext attack is discussed in section 4, and section 5 draws the conclusion.

## 2. Proposed Method

We propose to decompose the image into two parts, according to their interests, and encrypt each one with the same cryptosystem using a Dynamic Chaotic Look-Up Table [14]. It is then a hierarchical encryption. The key used in our cryptosystem is the same for encryption and decryption. Key $K$ is composed of three sub-chaotic cryptosystems ($SK_1$, $SK_2$ and $SK_3$). Every Sub-key is defined by the following combination SK = ($x_0$, $\mu_0$, $x_{0XOR}$, $\mu_{0XOR}$) where all the parameters are real numbers.

At first, we divide the image (Fig. 1-a) into two parts, according to their interests in the diagnosis. They are in the form of sharp changes in the intensity of the medical image. Only discontinuities and contours are present in the image. Although the rest of the image shows no abnormality, it can recognize the body. We then propose to use an expended edge detector "Canny" (Fig. 1-b) to separate between the homogeneous areas contours (Fig. 1-c). The expansion aims to enlarge the surface area of interest

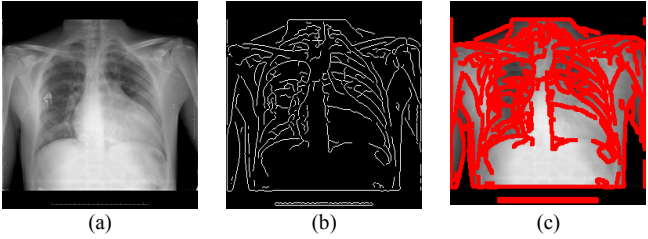

Fig. 1 Decomposition of the image by the dilated "Canny" detector: (a) Original image, (b) Detection of contour (using "Canny" detector), (c) Separate homogeneous areas contours (red).

The application of this separation is only the seven strongest plan of the original image. We can separate the plan with the lowest weight for its negligible impact on the image and its quality of clinical diagnosis. In fact, we will insert, in the lower level of our cryptogram, the detector response Canny dilated after being encrypted. We allow the watermarking at the decryption phase to recognize the two subsets of the image. In Fig. 2, we illustrate the hierarchical organization of our chaotic cryptosystem.

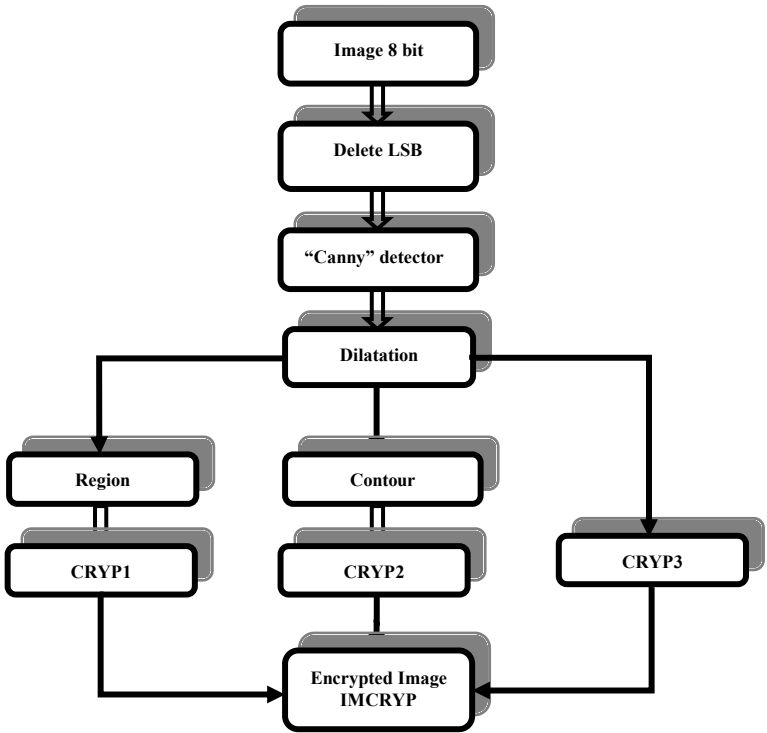

Fig. 2 Proposed encryption scheme.

When decrypting (Fig. 3), we perform an IMCRYP image divided into two subsets, the first is nothing but the bit plan 0 while the second subset involves the rest.

Decrypting the first bit plan which is designated by DECRYP3 helps us to obtain the dilated Canny contour. Based on this expanded deciphered contour, we subdivide the second subset into two parts: contour and region. The deciphering of these two parts gives us DECRYP2 DECRYP1. The fusion of DECRYP1 and DECRYP2, refers to the deciphered dilated Canny contour that will be inserted in the LSB plan of the resulting image which gives us the final decrypted image.

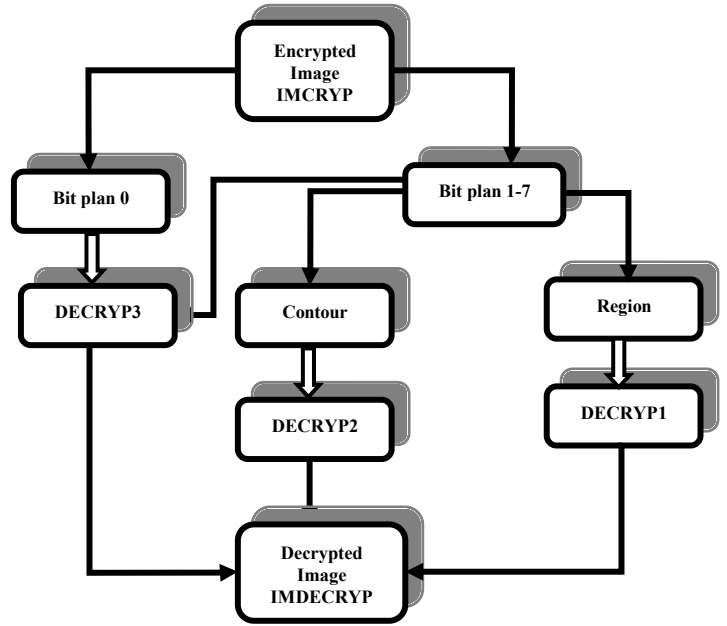

Fig. 3 Proposed decryption scheme.

## 3. Experimental results

In this proposed hierarchical encryption, a medical image 256x256 is used as shown in Fig. 4-a. The improved algorithm is implemented using MATLAB 7.6 on a personnel computer (PC) with a 1.67 GHz Intel Core 2 processor, 1 Go memory and 120 Go hard disk capacity. This mathematical tool encodes a real number of 8 bytes Thus, all the parameters are presented in 64-bit. Therefore, the proposed encryption image algorithm $\{(2^{64} \times 2^{64} \times 2^{64} \times 2^{64})^3 = (2^{256})^3\}$ has different combinations of the secret key.

Key $K$ hasthe following values : $\{SK_1 = (0.45, 3.801, 0.4003, 3.6701), SK_2 = (0.25, 3.8, 0.4, 3.67), SK_3 = (0.51, 3.805, 0.401, 3.77)\}$.

3.1 Histogram analysis

Fig. 4.a and 4.b show respectively the original image and the encrypted image by using $k_0$. The histogram of the plain image and the encrypted image are presented in Fig 4.c and 4.d.

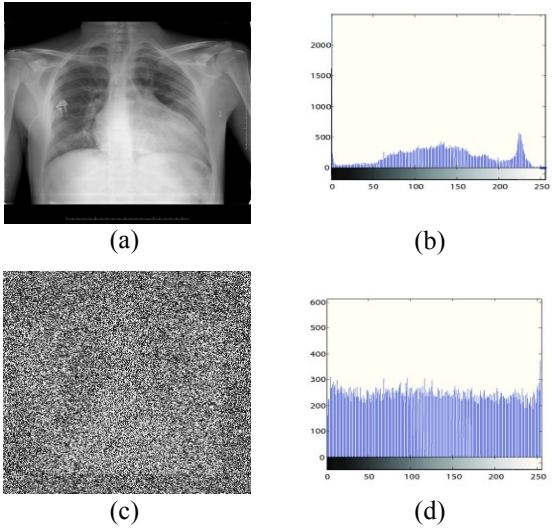

Fig. 4 Histogram analysis: (a) plain image, (b) plain image histogram, (c) encrypted image, (d) encrypted image histogram.

The results show that the histograms are different. We can clearly see that the plain image, Fig. 4.a, differs significantly from the corresponding encrypted one Fig. 4.b. Moreover, the histogram of the encrypted image is uniform, which makes it difficult to extract the pixels statistical nature of the plain image, Fig. 4.a, and makesdifficult any type of attack based on the analysis of the histogram of the encrypted image.

3.2 Differential analysis

To evaluate the performance of the proposed method, we use PSNR (Peak Signal to Noise Ratio) and UACI (Unified Average Changing Intensity) [15] to test the sensitivity of a single-bit change in the plain image.
The NPCR indicates the Number of Pixels Change Rates of the ciphered image while a pixel of the plain image is changed. NPCR is defined as:

$$NPCR = \frac{\sum_{i=0}^{M-1}\sum_{j=0}^{N-1} D(i,j)}{M \times N} \times 100 \quad (1)$$

D (i, j) is determined as follows:

$$D(i,j) = \begin{cases} 1 & if\ C_1(i,j) \neq C_2(i,j) \\ 0 & else \end{cases} \quad (2)$$

The Unified Average Changing Intensity (UACI) between these two images (plain and ciphered image) is defined by the following formula:

$$UACI = \frac{1}{M \times N}\sum_{i=0}^{M-1}\sum_{j=0}^{N-1} \frac{|C_1(i,j) - C_2(i,j)|}{255} \times 100 \quad (3)$$

The NPCR and UACI of the plain image shown in Fig 4.a are calculated and presented in Table 1. The objective of this analysis is to show that a small change in the image clearly introduces a major change in encrypted the image.

Table 1. Values results of NPCR and UACI for the plain image shown in fig. 4-a

| NPCR | UACI |
|---|---|
| 99.6720 | 32.3548 |

Given the results found after our test, we can conclude that our proposed method is resistant to differential attacks.

3.3. Correlation between two adjacent pixels

Another type of statistical analysis is the correlation coefficient analysis [16].

An image is often characterized by a strong correlation between a pixel and its neighboring pixels, in particular, the pixels on the same row or the same column or the same diagonal.

We calculate the correlation coefficient of a sequence of adjacent pixels using the following formula:

$$r_{xy} = \frac{cov(x,y)}{\sqrt{D(x)}\sqrt{D(y)}} \quad (4)$$

Here, x and y are the intensity values of two adjacent pixels in the image. $r_{xy}$ is the correlation coefficient. The cov(x,y), E(x) and D(x) are given as follows:

$$E(x) = \frac{1}{N}\sum_{i=1}^{N} x_i \quad (5)$$

$$D(x) = \frac{1}{N}\sum_{i=1}^{N}[x_i - E(x_i)]^2 \quad (6)$$

$$cov(x,y) = \frac{1}{N}\sum_{i=1}^{N}[(x_i - E(x_i))(y_i - E(y_i))] \quad (7)$$

N is the number of adjacent pixels selected from the image to calculate the correlation.

To test the correlation coefficient, we have chosen 2500 pairs of two adjacent pixels which are selected randomly from both the plain and encrypted image.

Figures 5, 6 and 7 illustrate the correlation of adjacent pixels in both the clear and the encrypted image. It is clear that visually neighboring pixels of a plain image are highly correlated, however this correlation decreases significantly in encrypted images.

We see that the adjacent pixels in the original image are highly correlated, while in the encrypted image there is no correlation. The distribution of this correlation of neighboring pixels of the encrypted image cannot prove that our cryptosystem allows the deduction of the values of neighboring pixels even with the knowledge of the value of a pixel.

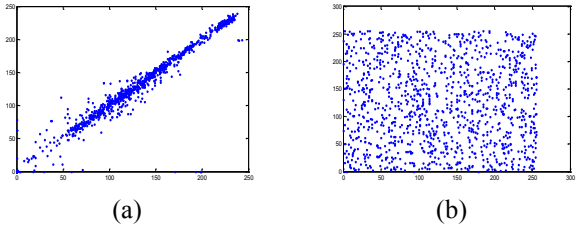

Fig. 5 Correlation between two diagonally adjacent pixels:
(a) in the plain image, (b) in the encrypted image.

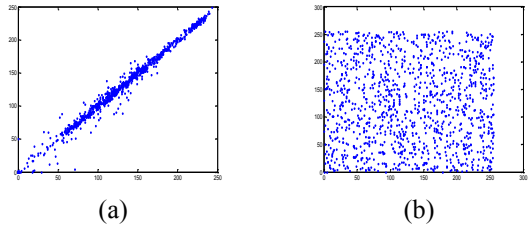

Fig. 6 Correlation between two horizontally adjacent pixels:
(a) in the plain image, (b) in the encrypted image.

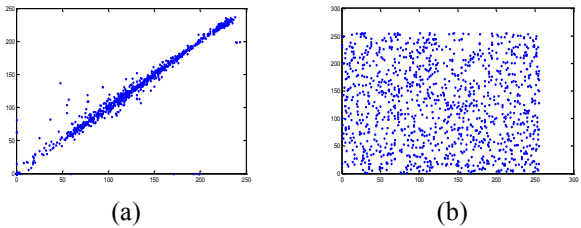

Fig. 7 Correlation between two vertically adjacent pixels:
(a) in the plain image, (b) in the encrypted image.

## 3.4 Key sensitivity

All cryptosystem should be very sensitive to every little change in the secret key [17]. In addition, we analyze the key sensitivity of our cryptosystem for encryption and decryption.

The secret key of our encryption algorithm is composed of four values $\{x_0, \mu_0, x_{0XOR}, \mu_{0XOR}\}$. These values are the parameters of the two Logistic Map functions used in our encryption algorithm.

We encrypt the original image shown in Fig 4.a with the secret key $k_0$ to analyze the sensitivity of our cryptosystem to some change in the secret key. Secondly, we propose to decrypt the cipher image encrypted with $k_0$ using keys with a difference by $10^{-15}$ on $x_0$, $\mu_0$, $x_{0XOR}$, $\mu_{0XOR}$.

Our cryptosystem is composed of three sub-cryptosystems. Therefore our key K is composed of three sub-keys $SK_1$, $SK_2$ and $SK_3$. To evaluate the key sensitivity due to the change in our hierarchical cryptosystem, we use three keys $K_1$, $K_2$ and $K_3$ that represent a slight modification respectively $SK_1$, $SK_2$ and $SK_3$ from K. As we kept the same structure for sub-cryptosystems, we introduce, in our tests, modifications ($10^{-10}$) on the value of $X_0$ for each sub-key (Table 2). The key K used has the following values : {0.45, 3.801, 0.4003, 3.6701, 0.25, 3.8, 0.4, 3.67, 0.51, 3.805, 0.401, 3.77}. We apply our hierarchical cryptosystem on the test image I with different keys K, $K_1$, $K_2$ and $K_3$ respectively for ICH, $ICH_1$, $ICH_2$ and $ICH_3$ images.

Table 2. Different keys used

|  | $K$ | $K_1$ | $K_2$ | $K_3$ |
|---|---|---|---|---|
| $SK_1(X_0)$ | 0.45 | $0.45 + 10^{-10}$ | 0.45 | 0.45 |
| $SK_2(X_0)$ | 0.25 | 0.25 | $0.25 + 10^{-10}$ | 0.25 |
| $SK_3(X_0)$ | 0.51 | 0.51 | 0.51 | $0.51 + 10^{-10}$ |

A visual inspection of the results of the changes in a sub-key cryptosystem gives hierarchical visually similar images. This effect is explained by the presence of a significant disorder in the encrypted images which we can not visually tell the difference between these images (Fig. 8). From these results, we can see that all the encrypted images are very different from the clear image.

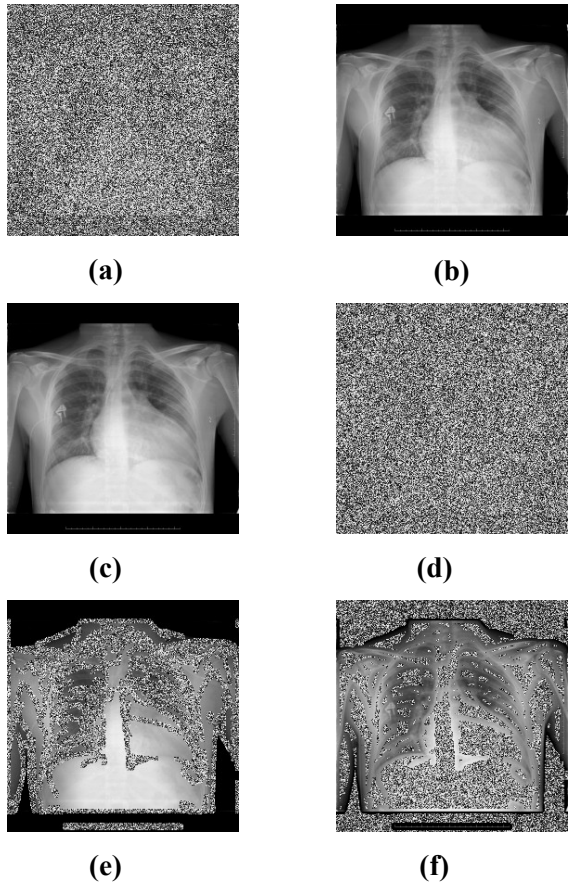

Fig. 8 Key sensitivity test: (a) plain image, (b) Encrypted image with $k_0$, (c) Decrypted image with k, (d) Decrypted image with $k_1$, (e) Decrypted image with $k_2$, (f) Decrypted image with $k_3$.

## 4. Cryptanalysis

In our study, we are based on the known plaintext attack to apply the stream key attack on our cryptosystem. This attack uses both the clear and the encrypted image to extract the decryption key and decrypt another encrypted image.

"Fig. 9" summarizes the key stream attack applied on our cryptosystem.

Fig. 9.c represents the key stream used to decrypt the encrypted image (Fig. 9.b). The result of this process is illustrated in Fig. 9.d. We note that the obtained image and the clear image (Fig. 9.a) are the same. After that, we use the key stream to decrypt the obtained image shown in Fig. 9.e, which represents the encrypted image of "thorax 2". Fig. 9.f proves the failure of the attack. Indeed, this last picture is quite different from the clear image "thorax 2" shown in Fig. 9.g.

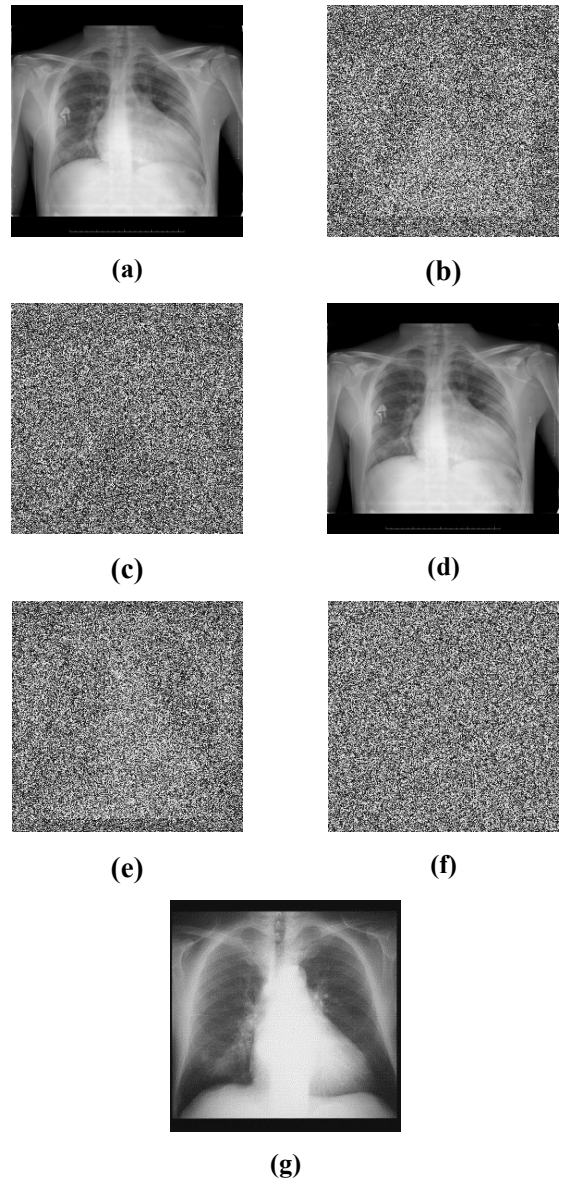

Fig 9. Failed crack attempt: (a) plain image "Torax 1", (b) encrypted image, (c) extracted key stream, (d) decrypted image "Torax 1", (e) encrypted image "Torax 2", (f) failed attempt to crack the cipher image of "Torax 2", (g) plain image "Torax 2".

The results in Fig. 9 show that our approach could not be broken by a known plaintext attack. This result is expected for the presence of a feedback in the cryptosystem.

## 5. Conclusions

In this paper, we proposed a new technique of hierarchical chaotic encryption. This technique is based on three chaotic cryptosystems. The first cryptosystem allows to partition the image pixels into two subsets. The two remaining sub-cryptosystems are used for the encryption of each subset. Thus, according to the key delivered to the recipient, we can define the rights of access to encryption of the received image. The evaluation of this technique has shown good performance against the cryptanalysis. If the results of sensitivity to the key changes seem unsatisfactory, this enhances the effect of access and hierarchical control of the cryptosystem.